\documentclass[a4paper,useAMS]{article} %{jarticle} %useAMS

\usepackage{upmath,amsfonts,amssymb,rotating,ai_fncs,eqn_center,cite}

\newcommand\bmath[1]{\ensuremath{\boldsymbol{#1}}}
\begin{document}
%\maketitle

\twocolumn[
\begin{center}
% Paper title
{\bf Molecular Dynamics Simulation of  Sputtering Process of Hydrogen  and Graphene Sheets}\\
~\\
% Authors
{\underline{\bf Hiroaki Nakamura}}$^{\dagger}$
%\thanks{$^\ast$Corresponding author. Email: nakamura@tcsc.nifs.ac.jp}$^{\ast\dagger}$, 
and {\bf Atushi Ito}$^\ddagger$, \\
% Affiliations
$^\dagger$National Institute for Fusion Science,
Oroshi-cho, Toki, Gifu 509-5292, JAPAN\\
$^\ddagger$Department of physics, Graduate School of Science, 
Nagoya University, Chikusa, Nagoya 464-8602, JAPAN
\end{center}
]

% Hiroaki Nakamura and Atsushi Ito
%7 pages (two column), 6 figures
\begin{sloppypar}
\begin{abstract}
{\bf
To clarify the yielding mechanism of small hydrocarbon molecules in  chemical sputtering between hydrogen and graphene sheets, we made  classical molecular dynamics simulation with modified Brenner's REBO potential which we proposed to deal with chemical reaction.
As the simulation model, we prepared more realistic physical system, which is composed of 160 incident hydrogen atoms and ten graphene multilayers, than our previous model.
From the present work, we found the following fact:
breaking the covalent bonds between carbon atoms by hydrogen  does not play an important role during destruction process of graphene structure, but  momentum transfer from incident hydrogen to graphene causes to destroy graphene structure.
Moreover, it is found that almost all fragments of graphene sheets form chain-shaped molecules, and that  yielded hydrocarbon molecules are composed of carbon chain and single hydrogen-atom. 
}

\noindent{\it Keywords}; graphene, graphite, hydrocarbon, sputtering, Molecular dynamics, Brenner potential, injection, hydrogen, undulation.

\end{abstract}

%%%%%%%%%%%%%%%%%%%%%%%%%%%%%%%%%%%%%%%%%%%%%% 
\section{Introduction}
Plasma-carbon interaction yields small hydrocarbon molecules on divertor region of a nuclear fusion device\cite{Nakano}.
Diffusing from divertor region to  core plasma region of fusion device, generated hydrocarbon takes  energy from the core plasma.
Therefore, the hydrocarbon is sometimes called ``dust" by the nuclear plasma researchers.
To remove an ill effect of the hydrocarbon on  core plasma, we need to know a creation mechanism of the hydrocarbons.
However, the creation mechanism of the hydrocarbons has not been elucidated yet.

As the first step to clarify the creation mechanism, we investigated, by computer simulation, collision process of  hydrogen atoms and one graphene sheet, which is regarded as one of   basic processes of  complex  plasma-carbon interaction in the previous work\cite{Ito}.
In the previous simulation,  we used   `classical' molecular dynamics (CMD) algorithm with  modified Brenner's REBO potential which we proposed to deal with chemical reaction between hydrogen and graphene\cite{Ito,rebo}.
From the previous work\cite{Ito}  in which an incident hydrogen kinetic energy $E_{\rm I}$ is less than 100 eV  to compare with experiments of divertor, it was found that an hydrogen-absorption rate of one graphene sheet depends on the incident hydrogen energy,  and  that the collision mechanism between a graphene and a hydrogen can be  classified into three types of processes:
absorption process for $E_{\rm I} < 5$ eV,  reflection process for $5$ eV $< E_{\rm I} < 50$ eV,  and  penetration process for $E_{\rm I} > 50$ eV.
Moreover, it was also found that when  hydrogen atom is absorbed by graphene,  the nearest carbon atom overhangs from the graphene which we called ``overhang structure".

Based on the above results, simulation model are extended  from a single  graphene sheet  to  multilayer graphene sheets in the present work. 
We adopt  three cases (5 eV, 15 eV and 100 eV)  as the incident energies of the hydrogen atoms $E_{\rm I}$. 
Each value corresponds to the absorption, the penetration, and the penetration process of the collision between single graphene sheet and hydrogen atoms, respectively.
Moreover, we use the same  simulation algorithm and the same interaction potential as the previous single-graphene sheet simulation.
The aim of the extension of simulation model  is that we reveal  a  more realistic sputtering process of graphene sheets and hydrogen atoms than the previous work.

In \S. \ref{sec.2}, we review the simulation method including the interaction potential which was proposed in the previous work.
We discuss the simulation results in \S. \ref{sec.3} and summarize a conclusion in the last section \S. \ref{sec.4}.
%%%%%%%%%%%%%%%%%%%%%%%%%%%%%%%%%%%%%%%%%%%%%%% 

\section{Simulation Method}\label{sec.2}
The simulation algorithm and the interaction potential are the almost same as the previous work\cite{Ito}.
%We review the outline of them here.
First, we review the interaction potential among hydrogen and graphite atoms.
The original Brenner's REBO potential\cite{rebo}, which was proposed on the basis of  past simulations\cite{Morse,Abell,Tersoff},  has the following form:
\begin{eqnarray}
U &=& \sum_{i,j(>i)} \sbk{V^R(r_{ij}) - \bar{b}_{ij}(\{\bmath{r}\}) V^A(r_{ij})}, \nonumber \\
  & & 
\end{eqnarray}
where $r_{ij}$ is the distance between atoms $i$ and $j$,  $V^A$ is an attractive term,  $V^R$ is a repulsive term and  the function $\bar{b}_{ij}(\{\bmath{r}\})$ includes all effects of molecular orbitals.
However, if chemical reaction occurs, the REBO potential breaks energy conservation.
Therefore, we proposed the following new functions expressing conjugation effects\cite{Ito}:
%\onecolumn
%\twocolumn[
\begin{eqnarray}
N_{ij}^{conj} &=& 1 + \sum_{k(\neq i,j)}^{carbon} f^c(r_{ik})C_N(N_{ki}^t) \nonumber \\
& & + \sum_{l(\neq j,i)}^{carbon} f^c(r_{jl})C_N(N_{lj}^t),
\end{eqnarray}
%\twocolumn
where $f^c$ is a cut-off function for a distance between atoms, and 
\begin{eqnarray}
C_N(x) &  &\nonumber \\
   &=& \left\{
              \begin{array}{cl}
				1 & {\rm if}\, x \leq 2,  \\
				\frac{1+\cos\left\{\pi (x - 2)\right\}}{2} & {\rm if}\, 2 < x \leq 3, \\
				0 & {\rm if}\, x > 3 ,  \\
              \end{array} \right.    \\
%\end{eqnarray}
%\begin{eqnarray}
	 N_{ki}^t &=& \sum_{j(\neq k,j)} f^c(r_{kj}) - f^c(r_{ki}) \nonumber \\
             &=& \sum_{j(\neq k,j,i)} f^c(r_{kj}) .
\end{eqnarray}
%Compared with the Brenner's original formulation\cite{rebo}, the second and the third terms of the function $N_{ij}^{conj}$ are not squared.
%The tricubic spline functions $F$ and $T$ in Ref. \cite{rebo} are redefined because they have $N_{ij}^{conj}$ as a variable.
%Because $F^{\prime}$ denotes a new function of $F$, $F^{\prime}(i,j,4) = F(i,j,6)$, $F^{\prime}(i,j,k \leq 5) = F(i,j,9)$  and $F^{\prime}(i,j,k) =  F(i,j,k)$ for the other cases.
%The spline $T(i,j,k)$ in Ref. \cite{rebo} was also redefined in a similar way\cite{Ito}.
%The above modifications of $N_{ij}^{conj}$, $F^{\prime}$ and $T^{\prime}$ derive differentiability at the cut-off point.
Modified Brenner's REBO potential conserves the total energy  during the chemical reaction.
%We executes the CMD simulation by the modified potential.

Next, we explain the simulation model in the present work.
We prepare ten graphene layers  each of which is composed by 160 carbon atoms.
All the layers are  perpendicular to $z$-axis  and have periodic boundary conditions for $x$ and $y$ directions. 
Distance between graphene sheets was set to $3.348$\AA \ in the initial state.
Graphite atoms at four corners of the first graphene sheet  from the top are fixed in all simulations.
In the second graphene sheet, the center graphene atom is fixed.  
After the third graphene sheet and later, the above fixed condition is repeated in all layers.
As the initial condition of the graphene layers, we prepare the equilibrium state of carbon atoms in all graphenes with a temperature $300$K. % by Nos\'{e}-Hoover thermostat\cite{Nose,Hoover}.
%After injection of the hydrogen atoms,  Nos\'{e}-Hoover thermostat is removed and carbon atoms move obeying Newton's laws with the interaction potentials among carbon and hydrogen atoms.

We shoot one incident hydrogen atom at a time every $5\times 10^{-14}$ seconds from $z=120$\AA \  plane to the first graphene which is located  at $z=3.348$\AA $\times (4+1/2) = 15.066$\AA, \ until the total number of the hydrogen atoms becomes 160.
We make the three simulations each incident hydrogen of which is set to 5 eV,  15 eV or 100 eV, respectively.
The $x$-$y$ coordinates of the incident hydrogen atom at $z=120$\AA \  plane are given as a pair of  random numbers ever time.
Each injection angle of the incident hydrogen atom between $z$-axis is set to a uniform random number from 0$^{\circ }$ to 60$^{\circ}$.

If an incident hydrogen atom is reflected by the graphenes and reaches $z=120$\AA \ plane, the reflected hydrogen atom is omitted from the simulation.
If the incident atom penetrates the graphene sheets and reaches $z=-120$\AA \ plane, the penetrated hydrogen atom is  also omitted from the simulation.

To integrate the equation of the motion, the second-order symplectic integration method is used in our simulation\cite{90Suzuki,93Umeno,05Hatano,90Yoshida}.

\section{Simulation Results and Discussions}\label{sec.3}
We simulated sputtering process of hydrogen atoms and graphene sheets with the three incident-energy cases, namely $E_{\rm I} =5$ eV, $E_{\rm I} =15$ eV and $E_{\rm I} =100$ eV.
Each incident energy corresponds to absorption process, reflection process or penetration process of sputtering between  a single graphene sheet and a hydrogen atom, respectively\cite{Ito}.
We show snapshots  of graphene multilayer structure and injected hydrogen atoms in the above three cases   in Figs. \ref{fig.E5eV}, \ref{fig.E15eV} and \ref{fig.E100eV}, respectively
We also plot the radial distribution functions of  carbon atoms for the first and the third sheets from the top in Figs. \ref{fig.E5eVgr}, \ref{fig.E15eVgr} and \ref{fig.E100eVgr}.
Detail of each simulation result is shown in the followings subsection.

%%%%%%%%%%%%%%%%%%%%%%%%%%%%%%%   5  e V     %%%%%%%%%%%%%%%%%%%%%%%%%%%%%%%%%%%
%%%%%%%%%%%%%%%%%%%%%%%%%%%%%%%   5  e V     %%%%%%%%%%%%%%%%%%%%%%%%%%%%%%%%%%%
%%%%%%%%%%%%%%%%%%%%%%%%%%%%%%%   5  e V     %%%%%%%%%%%%%%%%%%%%%%%%%%%%%%%%%%%
\subsection{The case of  $E_{\rm I}$ = 5 eV}
\begin{figure}[h]
\scalebox{1.5}{(a)}
\begin{center}
\includegraphics[width=0.4\textwidth, clip]{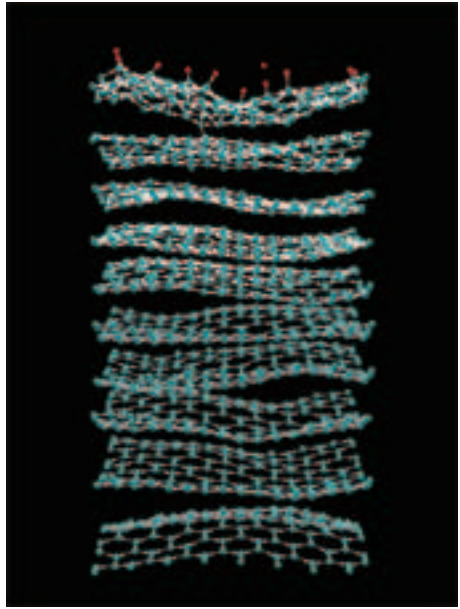}
\end{center}
%\vspace*{0.5cm}
\scalebox{1.5}{(b)}
\begin{center}
\includegraphics[width=0.4\textwidth, clip]{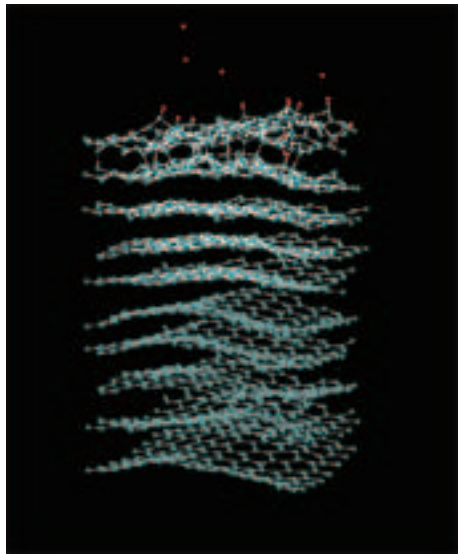}
\end{center}
\caption{\footnotesize \label{fig.E5eV}
Snapshots  of  graphene multilayer structure and injected hydrogen atoms in the case that $E_{\rm I}$=5 eV.
(a) About fifteen hydrogen atoms connect with the graphene sheet in the case that the total number of the injected hydrogen atoms $n_{\rm H}$ is 79. Each hydrogen atom and the surrounding carbon atoms compose the stable ``overhanged" state; 
(b)  $n_{\rm H}$ has increased to 159  and the first graphene sheet contacts with  the second one. The graphene structure is damaged. The increase in the number of the overhanged hydrogen atoms is not  as large as the increase in $n_{\rm h}$, {\it i.e.}, $159-79=80.$  The third sheet remains almost unchanged.}
\end{figure}

This situation corresponds to the absorption process between a hydrogen atom and a single graphene sheet\cite{Ito}.
It was intuitively expected that incident hydrogen atoms destroy the  graphene multilayer structure  from the surface graphene sheet  by  breaking covalent bonds between carbon atoms.  
The present simulation results shows that our prediction is partially correct and that the surface graphene, which is exposed to hydrogen atoms, is broken first.  
Contrary to our prediction, however, the leading cause of destruction of graphene sheet is not that the incident hydrogen breaks the covalent bond between carbon atoms, but that the incident hydrogen transfers its momentum to the carbon in the first graphene sheet (Fig.\ref{fig.E5eV}(a)).
The first graphene sheet, which gained momentum from the incident hydrogen, undulates more deeply, as hydrogen atoms are injected.
In course of time, amplitude of undulation becomes the same large as the distance of layers and the first graphene sheet contacts with the second one; then, the carbon in the first sheet interacts with the carbon in the second sheet (Fig.\ref{fig.E5eV}(b)). 
Thus, the first and the second sheets deform and their honeycomb-structures are broken.
The simulation in the next case of $E_{\rm I} = 15$ eV makes it clear that breaking covalent bonds between carbon by hydrogen do not play an important role in the above destruction mechanism.

\begin{figure}[h]
\scalebox{1.5}{(a)}
\begin{center}
\includegraphics[width=0.4\textwidth,clip]{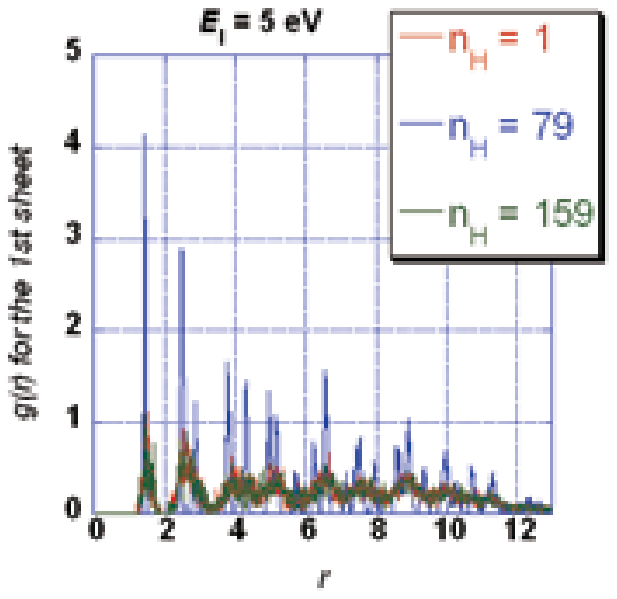}
\end{center}
%\vspace*{0.5cm}
\scalebox{1.5}{(b)}
\begin{center}
\includegraphics[width=0.4\textwidth,clip]{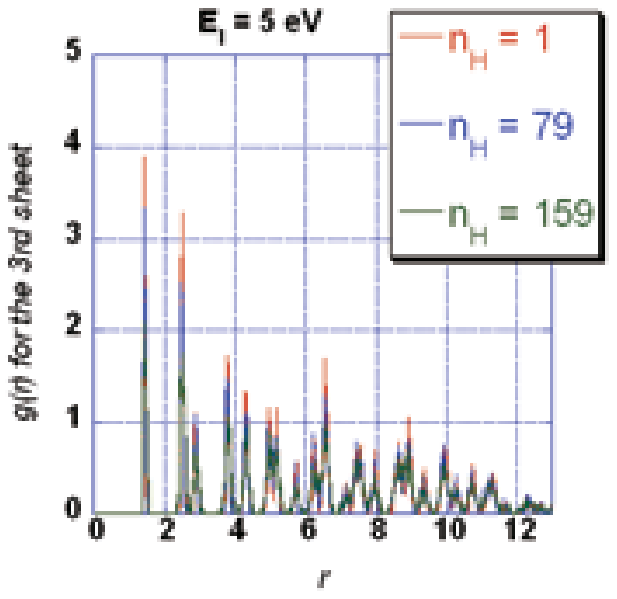}
\end{center}
\caption{\footnotesize \label{fig.E5eVgr}
Radial distribution functions $g(r)$ among carbon atoms in the first (a) and the third (b) graphene sheets for the three cases $n_{\rm H}$=1, $n_{\rm H}$=79 and   $n_{\rm H}$=159, where $n_{\rm H}$ denotes the total number of injected hydrogen atoms. The incident hydrogen energy is 5 eV.
}
\end{figure}

In Fig.\ref{fig.E5eVgr}(a), we plot the radial distribution functions $g(r)$ among carbon atoms in the first graphene sheet for the three cases $n_{\rm H}$=1, $n_{\rm H}$=79 and $n_{\rm H}$=159, where $n_{\rm H}$ denotes the total number of injected hydrogen atoms.
This figure shows that, as injected hydrogen atoms increases, the graphene structure of the first graphene sheet is breaking.
On the other hand, Fig.\ref{fig.E5eVgr}(b) shows that $g(r)$  among carbon atoms in the third graphene sheet remains almost unchanged during hydrogen injection, which denotes that  the third sheet is not destroyed by hydrogen.

%%%%%%%%%%%%%%%%%%%%%%%%%%%%%%%   15  e V     %%%%%%%%%%%%%%%%%%%%%%%%%%%%%%%%%%%
%%%%%%%%%%%%%%%%%%%%%%%%%%%%%%%   15  e V     %%%%%%%%%%%%%%%%%%%%%%%%%%%%%%%%%%%
%%%%%%%%%%%%%%%%%%%%%%%%%%%%%%%   15  e V     %%%%%%%%%%%%%%%%%%%%%%%%%%%%%%%%%%%
\subsection{The case of  $E_{\rm I}$ = 15 eV} 
\begin{figure}[ht]
\scalebox{1.5}{(a)}
\begin{center}
\includegraphics[width=0.4 \textwidth,clip]{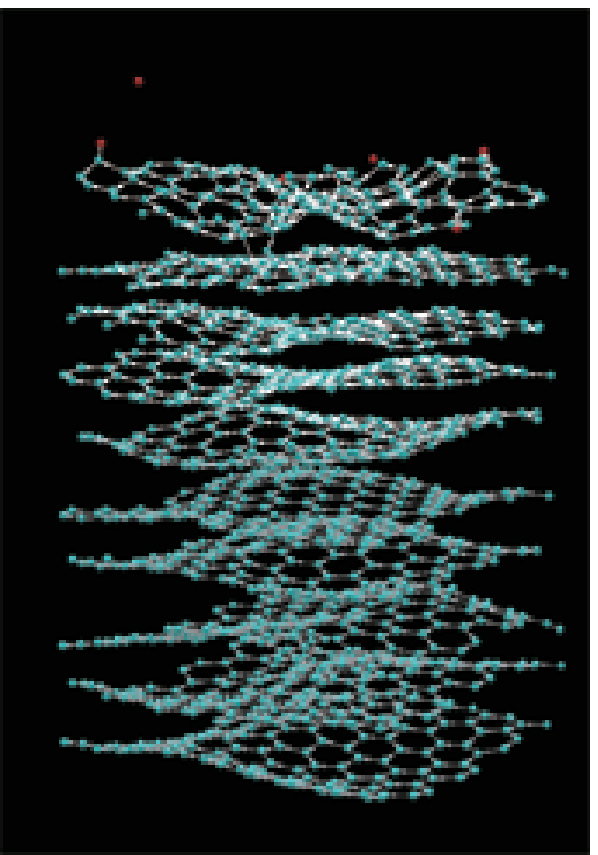}
\end{center}
%\vspace*{0.5cm}
\scalebox{1.5}{(b)}
\begin{center}
\includegraphics[width=0.4 \textwidth,clip]{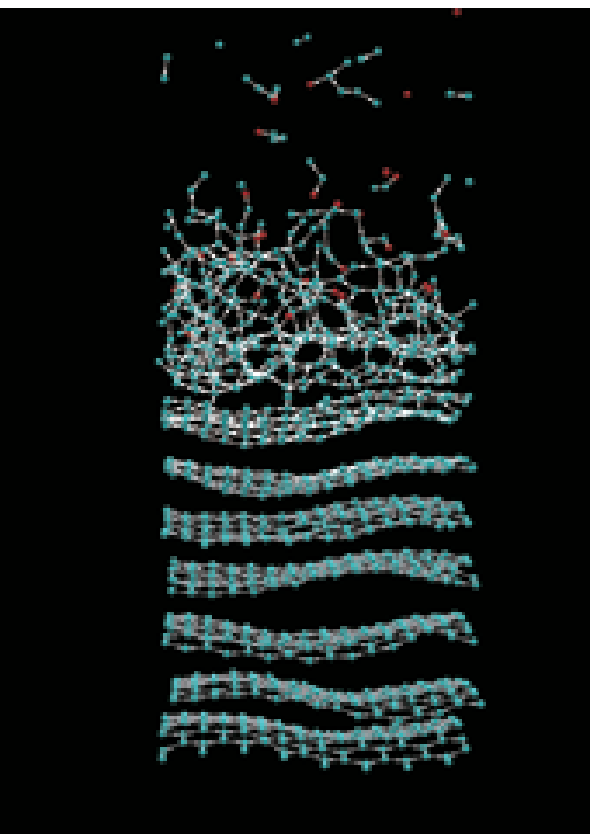}
\end{center}
\caption{\footnotesize \label{fig.E15eV}
Snapshots  of  graphene multilayer structure and injected hydrogen atoms in the case that $E_{\rm I}$=15 eV.
(a) Four hydrogen atoms connect with the graphene sheet in the case that the total number of the injected hydrogen atoms $n_{\rm H}$ is 30. Each hydrogen atom  and the surrounding carbon atoms  compose the stable ``overhanged" state; 
(b)  $n_{\rm H}$ has increased to 159  and the first graphene sheet contacts with  the second one. Thus, the graphene structure is damaged. 
Fragments of graphene sheets and hydrogen atoms form chain-shaped molecules.
The third sheet remains almost unchanged.}
\end{figure}

\begin{figure}[ht]
\scalebox{1.5}{(a)}
\begin{center}
\includegraphics[width=0.4\textwidth,clip]{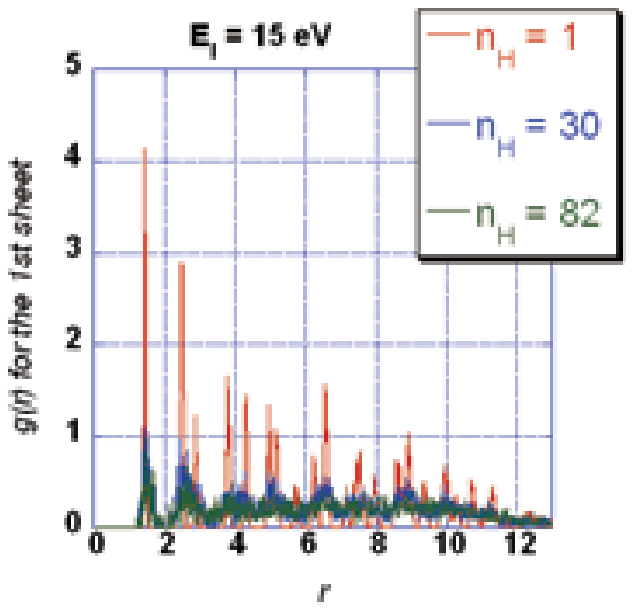}
\end{center}
%\vspace*{0.5cm}
\scalebox{1.5}{(b)}
\begin{center}
\includegraphics[width=0.4\textwidth,clip]{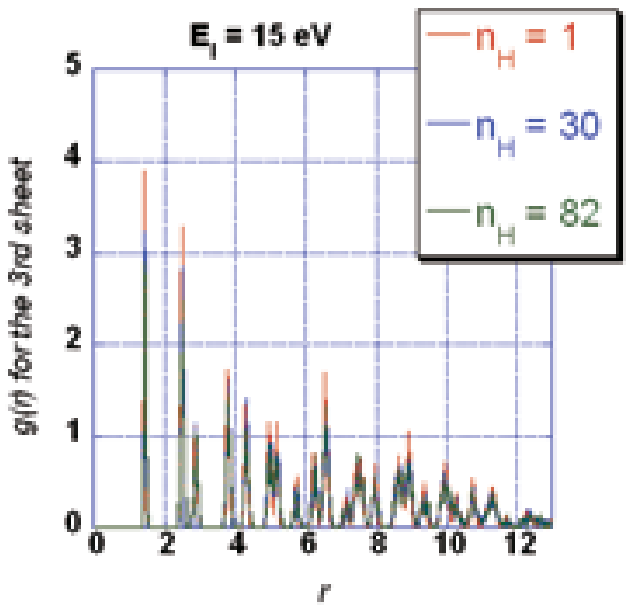}
\end{center}
\caption{\footnotesize \label{fig.E15eVgr}
The radial distribution functions $g(r)$ among carbon atoms in the first (a) and the third (b) graphene sheets for the three cases $n_{\rm H}$=1, $n_{\rm H}$=79 and   $n_{\rm H}$=159.
The incident hydrogen energy is 15 eV.
}
\end{figure}

The second situation corresponds to the reflection process of a hydrogen atom from a single graphene sheet\cite{Ito}.
Therefore, it is expected that it is  difficult for hydrogen to connect with graphene sheet.
This prediction is confirmed by Fig.\ref{fig.E15eV}(a), which shows that the number of overhanged hydrogen atoms is only four in spite of the fact that $n_{\rm H}$ = 30.
From this figure and Fig.\ref{fig.E15eVgr}(a), deformation of  the first graphene sheet is recognized in the case  that $n_{\rm H}$ = 30.
By the above fact and the simulation result in the case that $E_{\rm I}=5$ eV, we reached the following conclusion: breaking the covalent bonds between carbon atoms by hydrogen does not play an important role during  destruction process of graphene structure, but  momentum transfer from incident hydrogen to graphene causes to break graphene structure.

Furthermore, we obtained new information about  structures of hydrocarbon and carbon molecules which  are yielded by sputtering.
From Fig.\ref{fig.E15eV}(b),  it is found that almost all fragments of graphene sheets form chain-shaped molecules, and that yielded hydrocarbon molecules are composed of carbon chain and single hydrogen-atom.

%%%%%%%%%%%%%%%%%%%%%%%%%%%%%%%   100  e V     %%%%%%%%%%%%%%%%%%%%%%%%%%%%%%%%%%%
%%%%%%%%%%%%%%%%%%%%%%%%%%%%%%%   100  e V     %%%%%%%%%%%%%%%%%%%%%%%%%%%%%%%%%%%
%%%%%%%%%%%%%%%%%%%%%%%%%%%%%%%   100  e V     %%%%%%%%%%%%%%%%%%%%%%%%%%%%%%%%%%%
\subsection{The case of  $E_{\rm I}$ = 100 eV}
\begin{figure}
%\vspace*{0.5cm}
\scalebox{1.5}{(b)}
\begin{center}
\includegraphics[width=0.4\textwidth,clip]{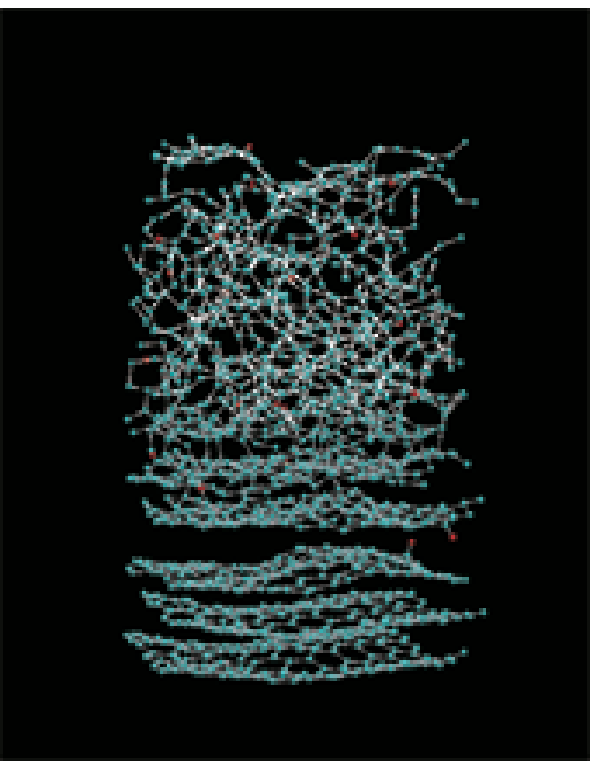}
\end{center}
%\vspace*{0.5cm}
\scalebox{1.5}{(c)}
\begin{center}
\includegraphics[width=0.4\textwidth,clip]{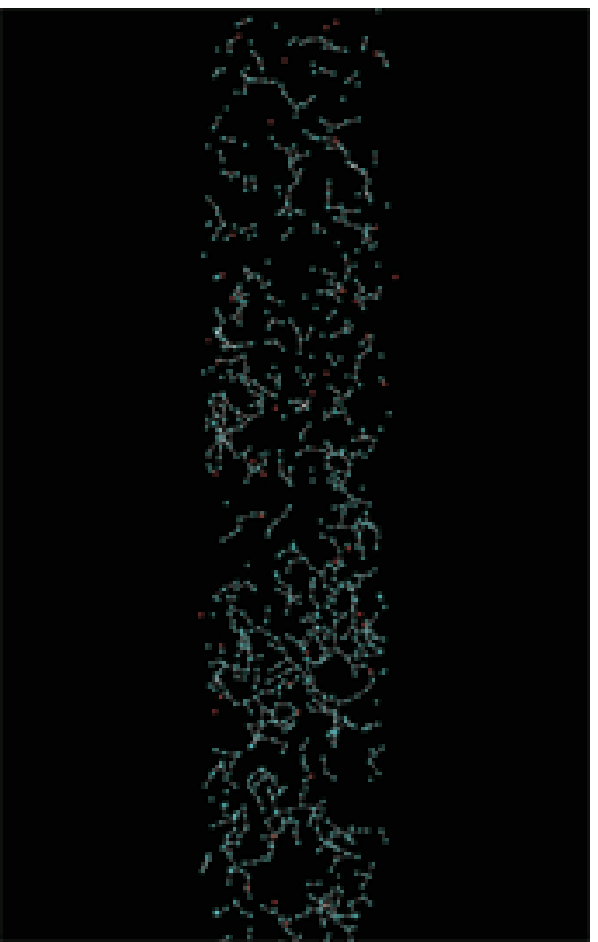}
\end{center}
\caption{\footnotesize \label{fig.E100eV}
Snapshots  of  graphene multilayer structure and injected hydrogen atoms in the case that $E_{\rm I}$=100 eV.
(a) After injected hydrogen atoms  penetrate  several graphene sheets, hydrogen atoms contact with the graphene sheet in the case that the total number of the injected hydrogen atoms $n_{\rm H}$ is 8. 
Graphene sheet is connected in both sides  with hydrogen. 
(b) $n_{\rm H}$ has increased to 30  and the upper five graphene sheet contact with  other sheets;
(c) graphene structure has already been broken to chain molecules when $n_{\rm h}$ becomes $159.$}
\end{figure}

\begin{figure}[h]
\scalebox{1.5}{(a)}
\begin{center}
\includegraphics[width=0.4\textwidth,clip]{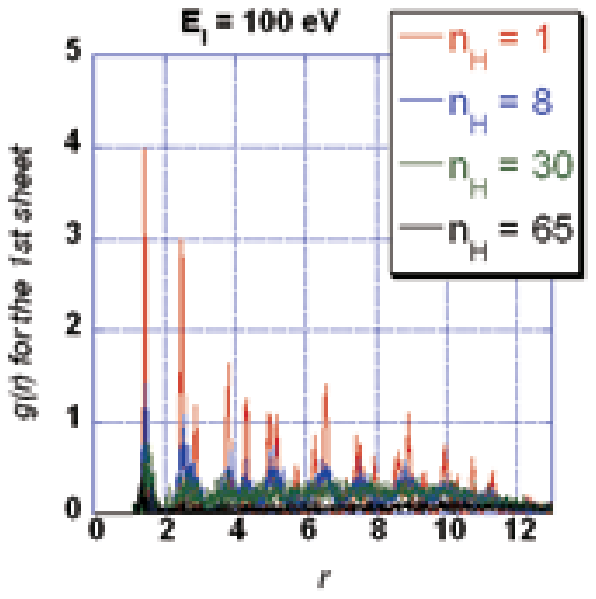}
\end{center}
%\vspace*{0.5cm}
\scalebox{1.5}{(b)}
\begin{center}
\includegraphics[width=0.4\textwidth,clip]{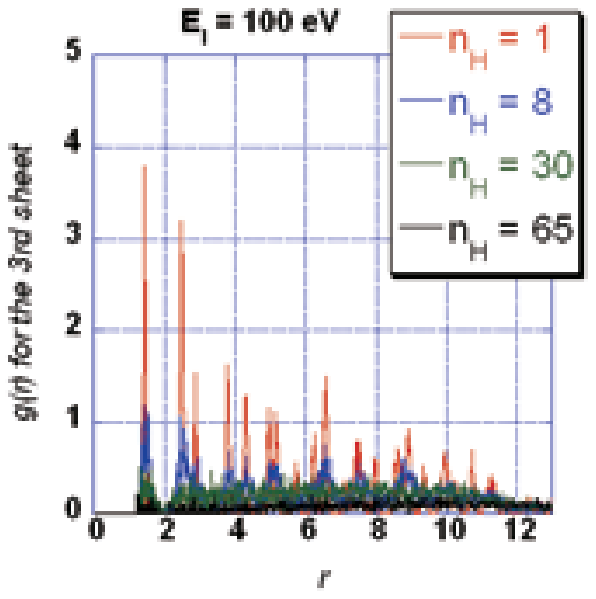}
\end{center}
\caption{\footnotesize \label{fig.E100eVgr}
Radial distribution functions $g(r)$ among carbon atoms in the first (a) and the third (b) graphene sheets for the three cases $n_{\rm H}$=1, $n_{\rm H}$=79 and   $n_{\rm H}$=159.
The incident hydrogen energy is 100 eV. 
The function $g(r)$ of the third sheet is similar to that of the first sheet.
}
\end{figure}

The last situation corresponds to the penetration process of a hydrogen atom from a single graphene sheet\cite{Ito}.
Injected hydrogen penetrates several graphene sheets with losing their kinetic energy, which is transferred to graphene sheets, until their energy becomes of the order of a few ten eV.
After that, the hydrogen atoms  begin to connect with carbon atoms (Fig.\ref{fig.E100eV}(a)).
From the standpoint of graphene sheets instead of hydrogen, only one side of graphene sheet  is exposed to hydrogen injection in the cases that $E_{\rm I} $=5 eV and $E_{\rm I} $=15 eV.
However, in the case of  $E_{\rm I}$ = 100 eV, both sides of graphene sheets are attacked by hydrogen atoms.
Therefore,  graphene structure is destroyed  rapidly  in a wide range (Fig.\ref{fig.E100eV}(b) and (c)).
Figure \ref{fig.E100eVgr} shows that the third sheet is destroyed in the almost same timing as the first sheet.

At the last stage of the simulation, all carbon molecules form chain-shaped structure.% instead of ring-shaped. 

\begin{figure}[h]
\scalebox{1.5}{(a)}
\begin{center}
\includegraphics[width=0.4\textwidth,clip]{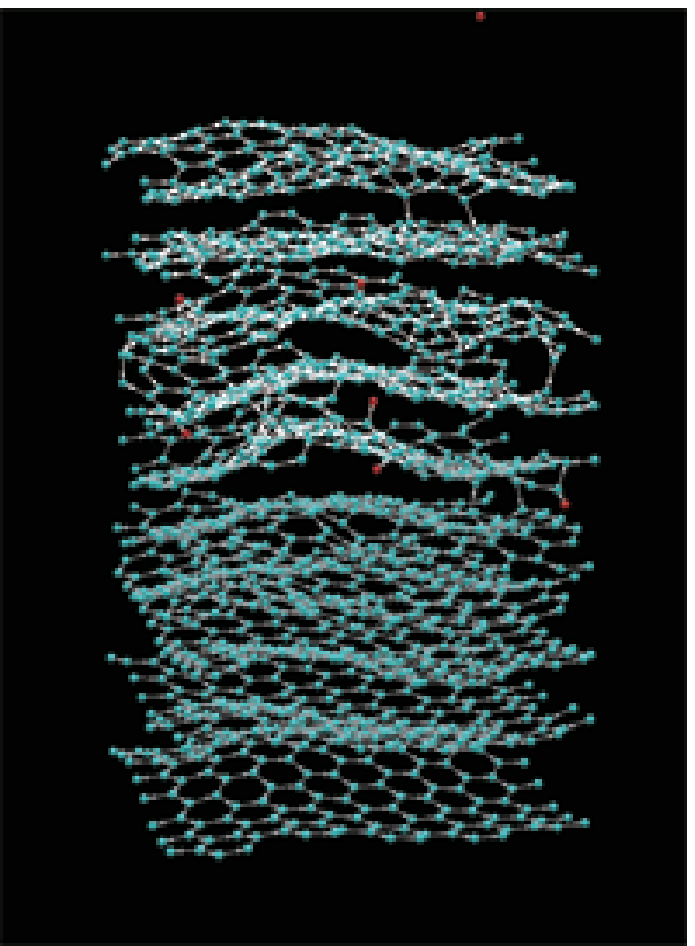}
\end{center}
\end{figure}

%%%%%%%%%%%%%%%%%%%%%%%%%%%%%%%   Conclusion  %%%%%%%%%%%%%%%%%%%%%%%%%%%%%%%%%%%
%%%%%%%%%%%%%%%%%%%%%%%%%%%%%%%   Conclusion  %%%%%%%%%%%%%%%%%%%%%%%%%%%%%%%%%%%
%%%%%%%%%%%%%%%%%%%%%%%%%%%%%%%   Conclusion  %%%%%%%%%%%%%%%%%%%%%%%%%%%%%%%%%%%
\section{Conclusion}\label{sec.4}
Destruction process of graphene multilayer due to hydrogen injection is demonstrated by CMD with the modified REBO potential.
From our simulation, we clarify the destruction process of graphene sheets as follows:
(1) hydrogen atoms transfer their momentum to graphene sheet; (2) the graphene is undulated and  touches  the neighboring graphene sheets; (3) inter-sheet interaction of carbon atoms works and  breaks graphene structure.
Moreover, it is found that almost all fragments of graphene sheets form chain-shaped molecules, and that   yielded hydrocarbon molecules are composed of carbon chain and single hydrogen-atom.

\section*{Acknowledgments}
The work is supported partly  by the National Institutes of Natural Sciences undertaking for Forming Bases for Interdisciplinary and International Research through Cooperation Across Fields of Study, and partly by Grand-in Aid for Exploratory Research (C), 2006, No.~17540384 from the Ministry of Education, Culture, Sports, Science and Technology. 

%%%%%%%%%%%%%%%%%%%%%%%%%%%%%%%%%%%%%%%%%%%

\end{sloppypar}

\begin{thebibliography}{9} 
\bibitem{Nakano}
     T. Nakano, H. Kudo, S. Higashijima, N. Asakura, H. Takenaga, T.  Sugie, and K. Itami,  ``Measurement of the chemical sputtering yields of ${\rm CH_4/CD_4}$ and ${\rm C_2H_x/C_2D_x}$ at the carbon divertor plates of JT-60U"  {\it Nucl. Fusion} {\bf 42}, 689-696 (2002).
\bibitem{Ito}
     A. Ito and H. Nakamura, ``Molecular Dynamics Simulation of Collision between a Graphite and a Hydrogen Atom",     {\it J. Plasma Phys.} in press. 	
\bibitem{rebo}
		 D. W. Brenner,  O. A. Shenderova,  J. A. Harrison, S. J. Stuart,  B. Ni and S. B.  Sinnott,``A second-generation reactive empirical bond order (REBO) potential energy 	expression  for hydrocarbons",{\it J. Phys. Condens. Matter} {\bf 14}, 783--802 (2002).
\bibitem{Morse}
	    P. M. Morse, ``Diatomic molecules according to the wave mechanics. II. vibrational levels", {\it Phys. Rev.} {\bf 34}, 57--64 (1929).
\bibitem{Abell}
		 G. C. Abell, ``Empirical chemical pseudopotential theory of molecular and metallic bonding",	{\it Phys. Rev. B} {\bf 31}, 6184--6196(1985).
\bibitem{Tersoff}
		J. Tersoff,	``{New empirical approach for the structure}", {\it Phys. Rev. B} {\bf 31}, 6184--6196(1985).
\bibitem{90Suzuki}
		M. Suzuki,	``Fractal decomposition of exponential operators with applications to many-body theories and Monte Carlo simulations", {\it Phys. Lett.} {\bf A 146}, 319--323 (1990).
\bibitem{93Umeno}
        K. Umeno and M. Suzuki,	``Symplectic and intermittent behaviour of Hamiltonian flow", {\it Phys. Lett.} {\bf A 181}, 387--392 (1993).
\bibitem{05Hatano}
N. Hatano and M. Suzuki, ``Finding Exponential Product Formulas of Higher Orders",  in {\it Quantum Annealing and related Optimization Methods}, Lecture Notes Phys. vol. 679, A. Das and B. K. Chakrabarti (Eds.), Springer, Heidelberg, (2005). Available on line at: arXiv.org:math-ph/0506007.
\bibitem{90Yoshida}
H. Yoshida, ``Construction of higher order symplectic integrators",{\it Phys. Lett.} {\bf A 150}, 262--268 (1990).
%\bibitem{Nose}
%		 S. Nos\'{e}, ``A unified formulation of the constant temperature molecular dynamics methods",	{\it J. Chem. Phys.} {\bf81}, 511--519(1984).
%\bibitem{Hoover}
%		 W. G. Hoover, ``Canonical dynamics: Equilibrium phase-space distributions",	{\it Phys. Rev.} {\bf A 31}, 1695--1697(1984).
\end{thebibliography}
\end{document}